\documentclass[twocolumn,showpacs]{revtex4}

\usepackage{graphicx}
\usepackage{amsmath}
\usepackage[dvips]{epsfig}

\makeatletter
\def\btt#1{\texttt{\@backslashchar#1}}
\DeclareRobustCommand\bblash{\btt{\@backslashchar}}
\makeatother
\input{tcilatex}

\begin{document}

\title[]{Raman frequency shift in oxygen functionalized carbon nanotubes }
\author{Z. X. Guo$^{1}$, J. W. Ding$^{1,2}$, Y. Xiao$^{1}$, and D. Y. Xing$^{2}$ }
\affiliation{$^{1}Department$ $of$ $physics,Xiangtan$ $University,$ $Xiangtan$ $411105,Hunan,China$\\
$^{2}National$ $Laboratory$ $of$ $Solid$ $State$ $Microstructures,Nanjing$ $%
University,$ $Nanjing$ $210093,China$\\
email: jwding@xtu.edu.cn}
\date{\today}

\begin{abstract}
In terms of lattice dynamics theory, we study the vibrational properties of
the oxygen-functionalized single wall carbon nanotubes (O-SWCNs). Due to the
C-O and O-O interactions, many degenerate phonon modes are split and even
some new phonon modes are obtained, different from the bare SWCNs. A
distinct Raman shift is found in both the radial breathing mode and G modes,
depending not only on the tube diameter and chirality but also on oxygen
coverage and adsorption configurations. With the oxygen coverage increasing,
interesting, a nonmonotonic up- and down-shift is observed in G modes, which
is contributed to the competition between the bond expansion and
contraction, there coexisting in the functionalized carbon nanotube.
\end{abstract}

\maketitle

Chemical functionalization of carbon nanotubes could offer new and promising
avenues to process and assemble tubes, add sensing capabilities, or tune
their electronic properties, which are the subject of intensive research.%
\cite{1,2} When the functional groups are chemically bonded to the nanotube
wall, the tube geometry can be significantly changed. As a result, the C-C
force constant and thus the vibrational properties of nanotube will be
largely modified. For example, an obvious upshift in the tangential mode
vibrations (so-called G modes) was observed experimentally in both
single-walled carbon nanotubes (SWCNs) and multi-walled carbon nanotubes
(MWCNs) in the acid treatment.\cite{3,4} For the physically doped nanotubes,
such a frequency shift was usually attributed to the variations in the C-C
bonds on nanotube, induced by the charge transfer.\cite{5,6,7,8} In the
chemically processed nanotubes, however, there exists much stronger
interaction between the functional groups and nanotubes, which is of
particular importance in the characterization of nanotube-based device. So
far, there were few, if any, works reported on\emph{\ }how functional groups
affect Raman shift in the nanotube. For nanotube-based device applications,
this is an outstanding issue and its clarification is greatly desirable.

An oxygen molecule ($O_{2}$) can be regarded as a simplest kind of
functional groups. It was reported that oxygen molecule is very reactive to
SWCNs with diameter below 1nm, and the oxidation can even occur at room
temperature,\cite{9,10} forming an oxygen-functionalized SWCN (O-SWCN). Such
oxidation can dramatically influence the nanotubes' electrical resistance,
thermoelectric power, and local density of states. The Raman shift and even
some new vibrational modes may be expected in such an O-SWCN, which may be
characterized by Raman spectroscopy. Therefore, the study of the oxygen
chemisorption effects on Raman modes can be very helpful to explore the
underlying mechanism of the Raman shift in the functionalized nanotubes.

As two types of the most important Raman modes, the radial breathing mode
(RBM) and G modes are widely used in the estimation of diameter distribution
of nanotubes.\cite{11,12} To assign the Raman peaks of\ an O-SWCN, the
frequency shift in these modes should be fully taken into account. In
addition, an O-SWCN can form several possible configurations, leading to
different electronic properties\cite{10}, which depend not only on the
adsorbed sites and coverage but also on whether the O-O bond breaking or not.%
\cite{9,10} It was reported that an SWCN with high adsorbed oxygen has a
higher electrical conductance than one with less adsorbed oxygen.\cite{13}
How about the influence of the adsorption configurations and coverage on the
vibrational properties of an O-SWCN? In the lattice dynamics, it is still an
open question and should be further explored.

In this paper, we study the vibrational properties of the O-SWCNs in terms
of lattice dynamics theory. Due to the C-O and O-O interactions, many
degenerate phonon modes are split and even some new phonon modes are
obtained, different from the bare SWCNs. A distinct Raman shift is found in
RBM and G modes, depending not only on the tube diameter and chirality but
also on oxygen coverage and adsorption configurations. With the oxygen
coverage increasing, interesting, a nonmonotonic up- and down-shift is
observed in G modes, which is contributed to the competition between the
bond expansion and contraction, there coexisting in the functionalized
carbon nanotube.

As typical example, we consider the cycloaddition structure for the $O_{2}$
adsorption on achiral tube. From previous report, the two configurations, $%
O_{2}$ on top of an axial C-C bond (site TA) and on top of a zigzag C-C bond
(site TZ), can lead to chemisorbed structures via cycloaddition (also see
Fig. 1), which shall be the focus of present work. To achieve the optimum
structure of the O-SWCNs, first-principle plane-wave pseudopotential density
functional theory (DFT) is used, performed in the CASTEP code.\cite{14} For
a simplification, we choose a unit cell of about 4.26\AA\ to be a supercell
in zigzag tube, while two unit cell of about 4.92\AA\ in armchair tube.\cite%
{15} The validity of the optimum results has been confirmed by using even
larger supercell.

Fig. 1 shows the optimized geometries of a (10,0) O-SWCN in both TA and TZ
configurations, which are similar with the previous results.\cite{9,10}
Interestingly, both the expansion and contraction of the C-C bonds are
obtained in the O-SWCN, different from the physically doped SWCNs only with
a bond expansion or contraction. In TA configuration, for example, three C-C
bonds at the adsorbed site are markedly expanded to be 1.500, 1.498 and 1.521%
\AA , respectively, while the bonds near the adsorbed site is contracted,
there appearing a minimal length of about 1.374\AA , shorter than 1.417\AA\ %
of the bare tube. Far from the adsorbed site, little change in the C-C bonds
is observed, almost not affected by the oxidation. The bond expansion can be
understood by the fact that a strong C-O covalent bond has been formed by
the introduction of holes into the $\pi ^{\ast }$ orbital,\cite{7} forming a
sp$^{3}$-like rehybridization at the adsorbed site.\emph{\ }This leads to
the bond length increasing, different from the Br$_{2}$ donor graphite
intercalation compounds, in which the C-C bonds still keep sp$^{2}$\
hybridization. As for the bond contraction, it is mainly contributed to the
charge transfer from SWCN to oxygen,\cite{5,6,7,8} similar to that in the
physically doped SWCNs, where the gain (removal) of electrons induces the
C-C bond expansion (contraction) and thus resulting in a downshift (upshift)
in G modes.\cite{7,8} The coexistence of the bond expansion and contraction
may indicate a new mechanism of Raman shift in the functionalized SWCNs.

For the lattice dynamic calculations, the vibrations of the carbon atoms on
the nanotube can be modeled by the force constant model in terms of the
force constants of graphene, up to the fourth next neighboring interaction.%
\cite{16} To incorporate the oxygen adsorption effects, the C-C force
constants have been corrected through the distorted bond lengths obtained
above.\cite{17,18} As for the C-O and O-O bond interactions, we use the
well-known Tersoff-Brenner bond order potential,\cite{19} and the potential
parameters in Ref.20. Shown in Fig. 2 are the phonon dispersions of the
(10,0) bared SWCN and O-SWCN in TA configuration. For the O-SWCN, there are
three acoustic phonon modes at around $\Gamma $ point: two transverse, one
longitudinal, similar with the bared SWCN. However, the twisting mode has a
non-zero frequency of about $20$cm$^{-1}$ at the center of Brillouin zone,
different from the bared SWCN, which is attributed to the tube geometry
distortion and\textit{\ }the oxygen vibrational interaction. Due to the tube
symmetry destruction, interestingly, many doubly degenerate phonon modes are
split. For example, the lowest E$_{2g}$\ mode is split into two absolute
subbands at 58cm$^{-1}$\ and 68cm$^{-1}$, shifted from 62cm$^{-1}$\ of the
bared SWCN. Such splitting will have an important influence on the
low-temperature specific heat.\cite{21} Also, some new phonon modes appear
below 300cm$^{-1}$ with nearly linear dispersion, which can be contributed
to the O-O bond stretching vibrations and the coupling vibrations between
the oxygen and the tube. Moreover, the highest frequency mode of 1607cm$%
^{-1} $ is newly obtained at $\Gamma $ point, higher than 1600cm$^{-1}$ of
the bare tube, which is ascribed to the strong C-O stretching vibrations.
Similar results are also obtained in TZ configuration. The splitting of
Raman modes and the new vibrational modes are expected to be observed by
Raman spectroscopic experiment.

Also shown in Table 1 are the RBM and G mode frequencies of the (10,0)
O-SWCN in both TA and TZ configurations. For the bare (10,0) tube, the RBM
frequency is obtained to be of 294cm$^{-1}$, in good agreement with the
experiment and the first principle results.\cite{25,26} In the TA (TZ)
configuration of the O-SWCN, the RBM frequency downshifts to be 278cm$^{-1}$%
(276cm$^{-1}$), decreased by 16(18)cm$^{-1}$, while the upshifts are
obtained in the G modes, increasing by 8(4)cm$^{-1}$ in E$_{1g}$, 6(13)cm$%
^{-1}$ in A$_{1g}$, and 7(3)cm$^{-1}$ in E$_{2g}$, respectively. It is shown
that the frequency shift in both RBM and G modes depends strongly on the
adsorption configurations, which should be considered in the nanotube
characterization by the Raman spectroscopic experiment.

\ To explore the dependence of the Raman shift on the tube diameter and
chirality, we calculate the RBM of armchair (5,5) and zigzag (8,0) O-SWCNs,
which have the similar tube diameter. For the (5,5) O-SWCN, the O-O bond
direction is perpendicular to the tube axis. In the case of the (8,0)
O-SWCN, the TA configuration is considered. For the bare and oxygenated
(8,0) SWCNs, the RBM frequencies are obtained to be 359 $cm^{-1}$and 329$%
cm^{-1}$, respectively. The frequency shift in RBM, $\Delta \omega
=-30cm^{-1}$ is much lager than that of (10,0) O-SWCN. This indicates the
dependence of the Raman shift on tube diameter, \textit{i.e.}, the larger
the frequency shift, the smaller the tube diameter. In the (5,5) O-SWCN, on
the other hand, the RBM frequency of about 272 $cm^{-1}$ shifts from 336$%
cm^{-1}$ of the bare (5,5) tube. The frequency shift ($\Delta \omega
=-64cm^{-1}$ ) is larger than two times that of the (8,0) O-SWCNs. At a
given tube diameter, obviously, an armchair tube is more sensitive to the
oxidation than a zigzag one, showing up a chirality-dependence of the Raman
shift. The results give a possibility to determine the tube diameter and
chirality of the O-SWCNs\emph{\ }by Raman spectroscopic experiment.

Now we further study the dependence of Raman shift on the coverage (defined
by O/C ratio $x$).\ As a typical example, Fig.3 shows the frequency of E$%
_{_{1}g}$ mode\ (one of G modes) in the (10,0) O-SWCN with TA configuration
as a function of $x$. For a given $x$, there may be much many
configurations. To obtain the qualitative relation between Raman shift and
the coverage, the oxygen atoms are simply considered to be averagely
adsorbed as possible, especially for a large $x$. From Fig. 3, E$_{_{1}g}$
mode first upshifts and then downshifts with $x$ increasing. There exists a
maximum of 1611cm$^{-1}$ at $x=0.25$, showing a nonmonotonic behavior of the
frequency shift. At half coverage ($x=0.5$), it even downshifts to be 1580cm$%
^{-1}$, lower than 1585cm$^{-1}$ of the bare SWCN. For other G modes, the
frequency shift in both A$_{_{1}g}$ and E$_{_{2}g}$ is also calculated,
which has the similar dependence on coverage to that in E$_{_{1}g}$. In
order to explore the origin of the G mode shift, we remove the C-O
vibrational\emph{\ }interactions and recalculate the E$_{_{1}g}$ mode in
terms of the same optimum geometry above. As a result, the frequencies are
overall downshifted, lower than those in the presence of the C-O vibrational
interactions, as shown in Fig. 3. Therefore, it is shown that there exists a
large contribution of the C-O vibration to the upshift in G modes.

In the absence of the C-O vibrational interactions, also, the nonmonotonic
behavior is still obtained for the frequency shift in E$_{_{1}g}$ modes.
This may be due to the competition between the rehybridization induced bond
expansion in the bonds at the adsorbed sites and the charge transfer induced
bond contraction in the bonds near the adsorbed site. At low coverage, the
bond contraction effect is dominant, leading to the upshift in G modes,
while the bond expansion effect would become predominant at high coverage,
resulting in an decrease of G mode frequency with $x$. Therefore, there may
exist a new competition mechanism between the bond expansion and
contraction, coexisting in the functionalized nanotube, different from the
physically doped SWCNs.\cite{7,22} In some alkali-metal doped nanotube
bundles, the similar changes in G modes with the coverage had been
experimentally observed via vapor phase doping or redox reaction.\cite{23,24}
Consequently, our results can contribute to the understanding of the origin
of the G mode shift in the functionalized carbon nanotube. In the case of
neglecting the C-O vibrational interactions, in addition, there exists a
locally minimal frequency of E$_{1g}$ mode at $x=0.1$. In this case, the
O-SWCN is considered to have its highest symmetry of mirror symmetry. To
understand this local minimum, we recalculate the E$_{1g}$ mode at the same
coverage but an asymmetrical configuration. The obtained frequency of 1592cm$%
^{-1}$ is higher than 1586cm$^{-1}$ in the symmetrical case, which further
shows the dependence of Raman shift on the configurations.

In summary, the vibrational properties of oxygen functionalized nanotubes
are studied in terms of lattice dynamics theory. Due to the C-O and O-O
vibrational interactions, many degenerate phonon modes are split and even
some new phonon modes are obtained, which can be observed in the Raman
spectroscopic experiment. Also, the Raman shift is observed in the O-SWCNs,
depending not only on the tube diameter and chirality but also on oxygen
coverage and adsorption configurations. It is found that the bond expansion
and contraction coexist in the functionalized carbon nanotube. With the
coverage increasing, a nonmonotonic behavior of the G mode shift is
obtained, which is contributed to the competition between the bond expansion
and contraction. The results can be helpful to the characterization and
practical application of functionalized carbon nanotube-based device.

{\large Acknowledgments:} This work was supported by National Natural
Science Foundation of China (Nos.: 10674113 and 10374046), Hunan Provincial
Natural Science Foundation of China (No.: 06JJ5006), and partially by
Scientific Research Fund of Hunan Provincial Education Department (No.:
06A071). 
\newline
\newline
\newline
\newline

\newpage
 
Table. 1. Frequencies of RBM and G modes of the bare (10,0) SWCN
and O-SWCN in both TA and TZ configurations (in units of cm$^{-1}$).

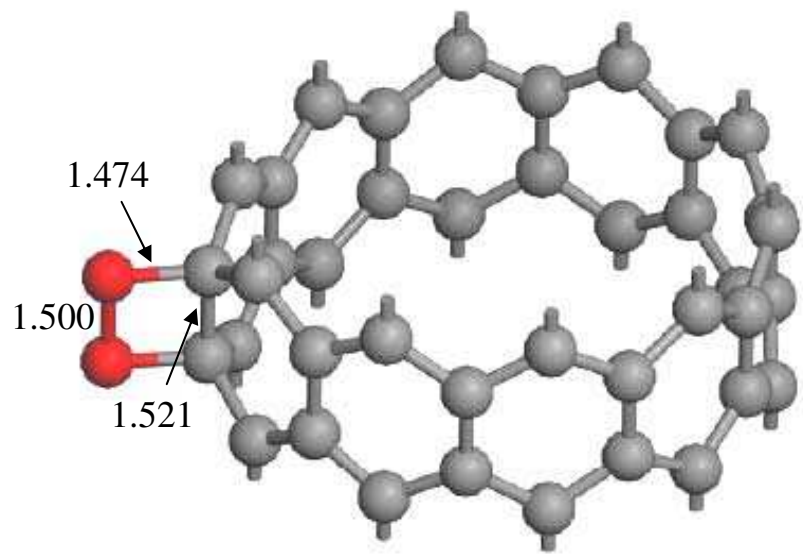
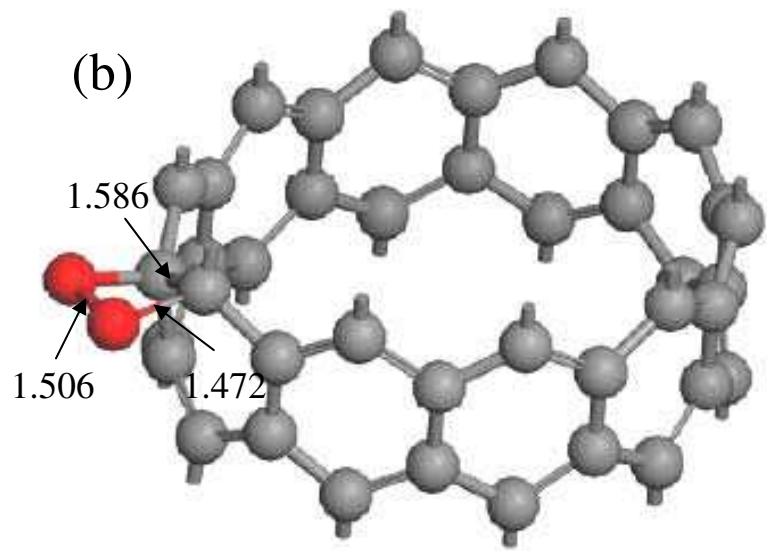
Figure. 1. Optimized structures of (10,0) O-SWCNs in (a) TA and (b) TZ
configurations.

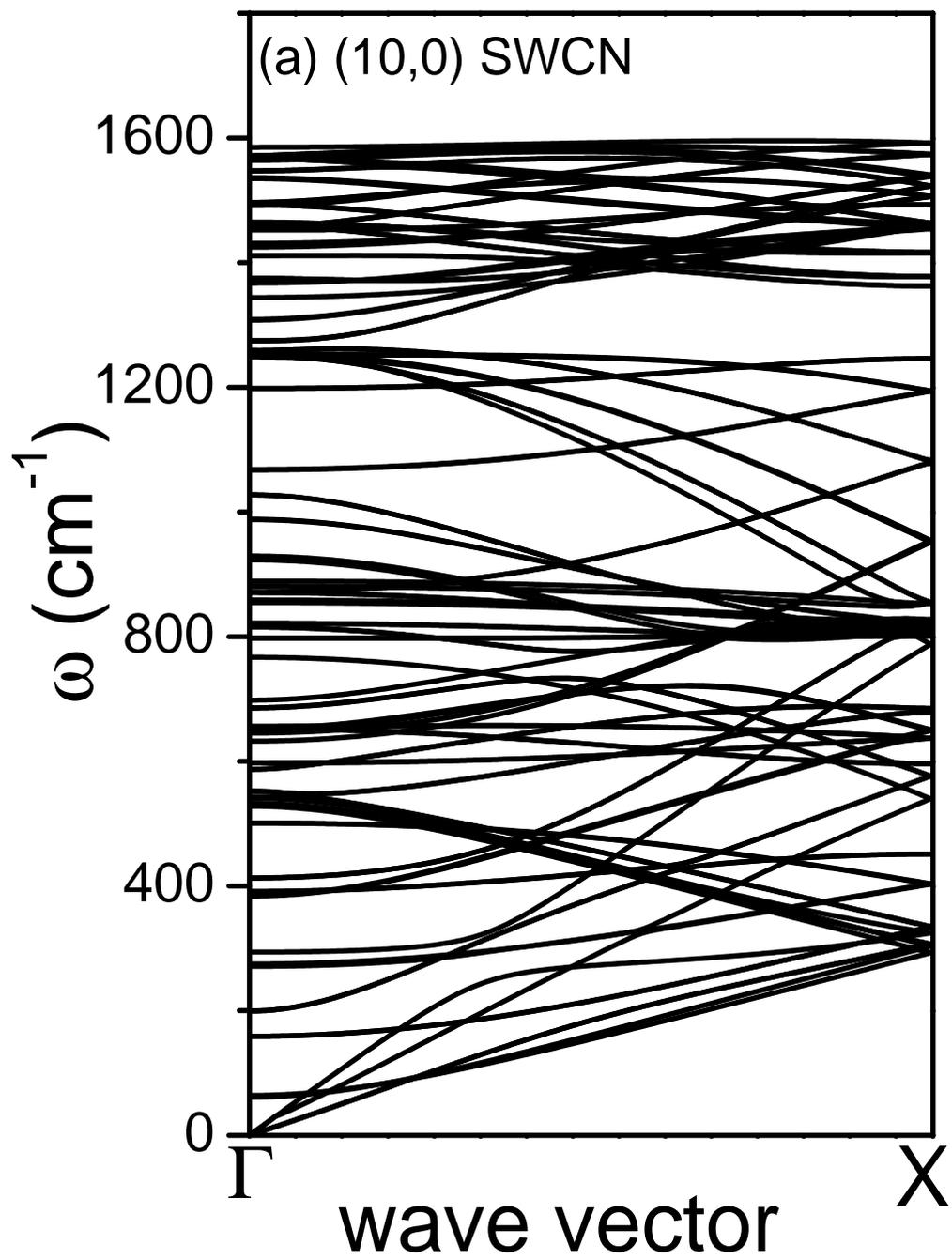
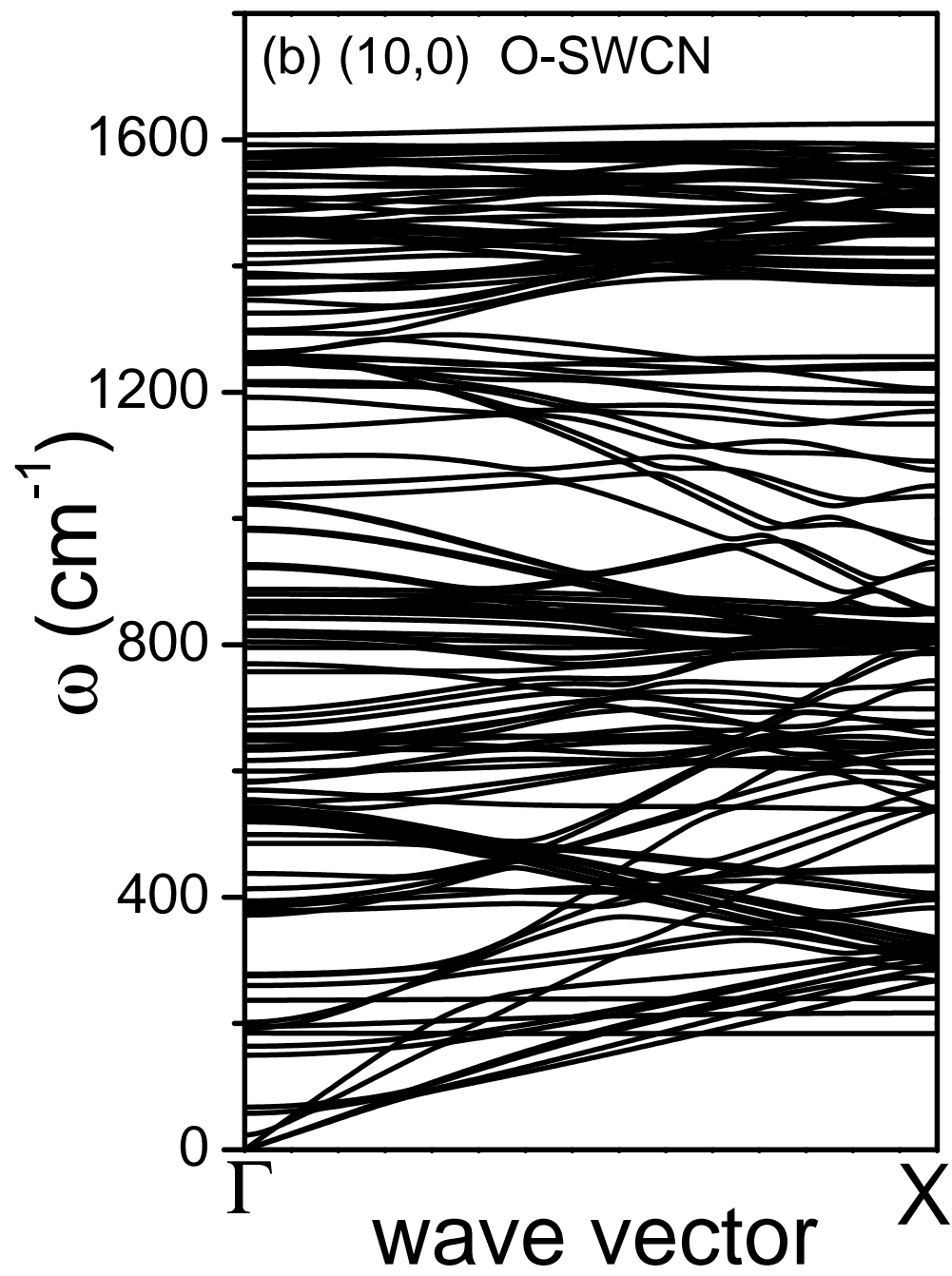
Figure. 2. Phonon dispersion relations of the (10,0) SWCN (a) and O-SWCN in
TA configuration (b).

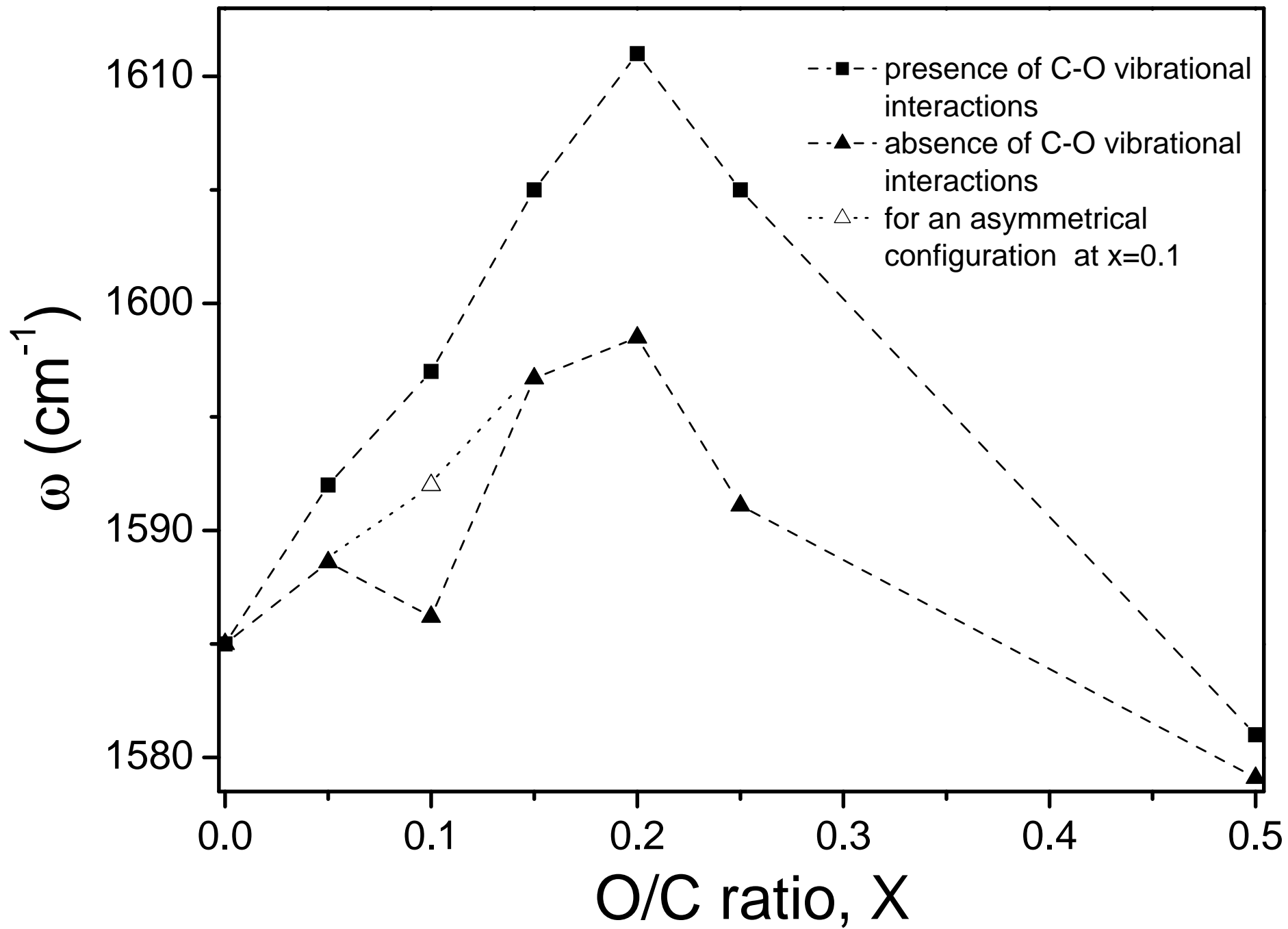
Figure. 3. E$_{1g}$ mode frequency of the (10,0) O-SWNT in TA configuration
as a function of the coverage x in both the absence and the presence of the
C-O vibrational interactions.

\end{document}